\documentclass[aps,prd,twocolumn,nofootinbib,superscriptaddress]{revtex4-2}

\pdfoutput=1

\usepackage[utf8]{inputenc} 
\usepackage{amsmath}
\usepackage{amssymb}
\usepackage{xcolor}
\usepackage{graphicx}
\usepackage{dcolumn}
\usepackage{bm}
\usepackage{microtype}
\usepackage{threeparttable}
\usepackage{mathtools}
\usepackage[colorlinks=true]{hyperref}
\usepackage{physics}
\usepackage{comment}
\usepackage[normalem]{ulem}

\newcommand{\stkout}[1]{\ifmmode\text{\sout{\ensuremath{#1}}}\else\sout{#1}\fi}

\newcommand{\be}{\begin{eqnarray}}
\newcommand{\ee}{\end{eqnarray}}
\newcommand{\ps}[1]{\bar{#1}}
\newcommand{\AH}[1]{{#1}}
\newcommand{\psf}[1]{\bar{#1}}
\newcommand{\AHf}[1]{{\tilde{#1}}}

\begin{document}

\title{Adiabatic hydrodynamization and quasinormal modes of nonthermal attractors}

\author{Matisse De Lescluze}
\email{matisse.delescluze@ugent.be}
\affiliation{Department of Physics and Astronomy, Ghent University, 9000 Ghent, Belgium}

\author{Michal P. Heller}
\email{michal.p.heller@ugent.be}
\affiliation{Department of Physics and Astronomy, Ghent University, 9000 Ghent, Belgium}
\affiliation{Institute of Theoretical Physics and Mark Kac Center for Complex
Systems Research, Jagiellonian University, 30-348 Cracow, Poland}

\author{Aleksas Mazeliauskas}
\email{a.mazeliauskas@thphys.uni-heidelberg.de}
\affiliation{Institute for Theoretical Physics, University of Heidelberg, 69120 Heidelberg, Germany}

\author{Bruno Scheihing-Hitschfeld}
\email{bscheihi@kitp.ucsb.edu}
\affiliation{Kavli Institute for Theoretical Physics, University of California, Santa Barbara,
California 93106, USA}

\author{Clemens Werthmann}
\email{clemens.werthmann@ugent.be}
\affiliation{Department of Physics and Astronomy, Ghent University, 9000 Ghent, Belgium}

\begin{abstract}
Nonthermal attractors govern the emergent self-similar dynamics of far-from-equilibrium quantum systems, from ultrarelativistic nuclear collisions to cold-atom experiments. Within the framework of adiabatic hydrodynamization, the approach to a nonthermal attractor is described by the decay of excited states of an effective Hamiltonian. Using an exactly solvable kinetic theory—the longitudinally expanding, overoccupied gluon plasma dominated by small-angle elastic scattering—we establish a direct correspondence between the eigenmodes of adiabatic hydrodynamization and the quasinormal mode spectrum of the nonthermal attractor. This equivalence suggests a general framework for identifying universal dynamical structures in nonequilibrium systems. As a byproduct, we derive analytic prescaling solutions for strongly longitudinally expanding systems.

\end{abstract}

\maketitle

\noindent \textbf{\emph{Introduction.--}} Understanding out-of-equilibrium dynamics and consequent thermalization of quantum many-body systems remains an important subject of contemporary research. A key goal is identifying organizing principles that classify the resulting emergent dynamics. Prime examples of universal out-of-equilibrium phenomena are nonthermal~\cite{Berges:2008wm,Berges:2014bba,Boguslavski:2018phs,Mikheev:2023juq} and hydrodynamic~\cite{Heller:2015dha,Soloviev:2021lhs,Jankowski:2023fdz} attractors, which are characterized by ``memory loss'': a quick collapse of time evolution curves onto a subspace in state space. How a system approaches this attractor behaviour is actively investigated topic~\cite{Tanji:2017suk,Chantesana:2018qsb,Mazeliauskas:2018yef,Schmied:2018upn,Mikheev:2022fdl,Du:2022bel,Heller:2020anv,Boguslavski:2023jvg,BarreraCabodevila:2025vir}, in particular by employing the Adiabatic Hydrodynamization (AH)~\cite{Brewer:2019oha,Brewer:2022vkq,Rajagopal:2024lou,Rajagopal:2025nca} and Quasinormal Mode (QNM)~\cite{DeLescluze:2025jqx} frameworks. In this Letter, we establish a correspondence between these two approaches in the context of a longitudinally expanding, overoccupied gluon plasma, modeling the earliest stages of high-energy nuclear collisions.

Nonthermal attractors or Nonthermal Fixed Points (NTFPs) correspond to attractive regimes of universal self-similar time evolution far-from-equilibrium~\cite{Berges:2008wm}. They appear in many different branches of physics: in cold atom experiments~\cite{PineiroOrioli:2015cpb,Mikheev:2018adp,Prufer:2018hto,Erne:2018gmz,Glidden:2020qmu,Huh:2023xso,Gazo:2023exc,Lannig:2023fzf,Martirosyan:2023mml}, in the study of the early Universe~\cite{Micha:2002ey}, and in ultra-relativistic heavy-ion collisions~\cite{Schlichting:2012es,Kurkela:2012hp,Berges:2013eia,Berges:2013fga,Berges:2013lsa,AbraaoYork:2014hbk,Kurkela:2015qoa,Schlichting:2019abc,Berges:2020fwq}. For the latter, at sufficiently high energies, the system is described by a gluon distribution function $f(\tau,p_z,p_\perp)$ that depends on proper time ($\tau$), longitudinal ($p_z$) and transverse momenta ($p_\perp$)~\cite{Schlichting:2019abc,Berges:2020fwq}. While in other cases the medium can be treated as isotropic, in heavy-ion collisions, a strong longitudinal expansion causes the distribution to be highly anisotropic at early times.
A NTFP is defined by the distribution function taking the form
\begin{equation}
    f(\tau, p_z,p_\perp)=\big(\tfrac{\tau}{\tau_\text{ref}}\big)^\alpha f_s\Big(\big(\tfrac{\tau}{\tau_\text{ref}}\big)^\gamma p_z,\big(\tfrac{\tau}{\tau_\text{ref}}\big)^\beta p_\perp\Big) \, ,
\label{eq: ideal NTFP}
\end{equation}
where $\tau_{\rm ref}$ is some reference time and $f_s$ is the scaling function. When $f$ takes on this form, it evolves self-similarly because over a long time range the state at a later time can be found by rescaling the amplitude and momenta by powers of time. The scaling function $f_s$ together with the scaling exponents $(\alpha,\beta,\gamma)$ characterize the NTFP. This characterization also applies to cold atoms experiments~\cite{PineiroOrioli:2015cpb,Mikheev:2018adp,Prufer:2018hto,Erne:2018gmz,Glidden:2020qmu,Huh:2023xso,Gazo:2023exc,Lannig:2023fzf,Martirosyan:2023mml} and early-Universe cosmology~\cite{Micha:2002ey}, albeit with only two scaling exponents $(\alpha,\beta)$ due to isotropy.

We consider a system following the Fokker-Planck-type Boltzmann equation~\cite{Mueller:1999pi,Blaizot:2013lga}
\begin{equation}
    \frac{\partial f}{\partial \tau} - \frac{p_z}{\tau} \frac{\partial f}{\partial p_z} = \lambda_0 I[f] \frac{\partial^2 f}{\partial p_z^2} \, ,
    \label{eq: FP Bjorken} 
\end{equation}
with $I=\tfrac{1}{\left(2\pi\right)^3}\int d^3p\, f^2$ and $\lambda_0$ a dimensionless parameter characterizing the coupling strength of the theory. This equation is motivated by the dominance of longitudinal dynamics over any transverse dynamics, allowing to keep only derivatives in $p_z$, and by overoccupation $f\gg 1$, allowing to neglect drag in favor of diffusion. With these simplifying assumptions, the system becomes effectively one-dimensional and admits a well-defined scaling solution. In fact, this model is \textit{exactly solvable} down to a single ordinary differential equation, as was first shown by~\cite{Brewer:2022vkq} by means of the AH framework. The exact NTFP solution is advantageous for establishing an explicit correspondence between AH and QNMs. In more general kinetic theories, such a solution will only be approximate such that the QNMs as defined here are approximate solutions. This model reproduces the scaling exponents $(\alpha,\beta,\gamma) = (-2/3,0,1/3)$ predicted by the canonical ``bottom-up'' thermalization scenario in heavy-ion collisions, which was first described in QCD effective kinetic theory (EKT)
~\cite{Baier:2000sb} (see also~\cite{Schlichting:2019abc,Berges:2020fwq}) and later confirmed by classical-statistical simulations~\cite{Berges:2013fga,Berges:2013eia}.


We study two frameworks describing the relaxation to a NTFP. Adiabatic hydrodynamization provides a global description of the dynamics in terms of instantaneous eigenstates of a time-dependent effective Hamiltonian, organizing the evolution as a hierarchy of modes. The construction relies on the existence of an adiabatic frame and, in practice, requires that the nonlinear dynamics can be cast into a form where such an effective Hamiltonian is well-defined and its spectrum is tractable. As a result, AH is most naturally applicable to systems with sufficiently simple effective Hamiltonians~\cite{Brewer:2019oha,Brewer:2022vkq,Rajagopal:2024lou,Rajagopal:2025nca}.

In contrast, quasinormal modes characterize the decay of perturbations around a reference state through a linearized analysis, yielding a universal spectrum of relaxation exponents. Near equilibrium, the notion of QNMs is widespread in usage. In particular, in holographic systems black hole QNMs dictate how a small perturbation to the equilibrium state relaxes back to equilibrium~\cite{Horowitz:1999jd,Kovtun:2005ev,Berti:2009kk}. Their analogs in weakly coupled systems are determined by the linearized time evolution operator derived from the Boltzmann equation. This concept has recently been extended to non-expanding, isotropic NTFPs \cite{DeLescluze:2025jqx}, where they take the form of power-law corrections. Since the QNM analysis is based solely on linearizing the evolution equations around an exact, self-similar attractor solution, it provides a general and systematic characterization of perturbations near this state, independent of the complexity of the underlying nonlinear dynamics. In this sense, QNMs provide a universal near-attractor description, while AH offers a more global but less generally applicable framework. 

As both the AH and QNM frameworks describe the same physical process, 
there should exist a correspondence between AH and QNMs of NTFPs 
near Nonthermal Fixed Points. More generally, such a correspondence can be expected in systems that (i) exhibit a self-similar attractor and (ii) admit a well-defined linearized evolution operator around it. Near such attractors, the dominant time dependence can be absorbed into rescaled variables, rendering the dynamics effectively stationary. In such a frame, an adiabatic description with a hierarchy of slowly decaying modes naturally emerges. Thus, adiabaticity is not an independent assumption but a consequence of the underlying self-similarity. We will demonstrate at both linear and nonlinear levels that the AH eigenmode decomposition can be expressed in terms of the NTFP plus QNMs. While our derivation is model-specific, it provides concrete evidence that such a correspondence arises whenever the two criteria stated above are fulfilled. 


\noindent \textbf{\emph{Adiabatic Hydrodynamization.--}} The AH framework~\cite{Brewer:2019oha,Brewer:2022vkq,Brewer:2022ifw,Rajagopal:2024lou,Rajagopal:2025nca} provides an organizing principle to understand the process of memory loss as a system 
hydrodynamizes starting from an out-of-equilibrium initial condition. It is based on mapping the Boltzmann equation into a Schr\"odinger-like equation $\partial_t f = - \mathcal{H}(t) f$, which allows to identify the eigenstates with lowest real parts of the corresponding eigenvalues as dynamically preferred, as they are the least decaying. The emergence of universal dynamics is then understood as a consequence of the presence of a gap in the eigenvalue spectrum of $\mathcal{H}$.

The organizing principle is \textit{adiabaticity}: instantaneous eigenstates smoothly transition into each other in time. While adiabaticity is not guaranteed a priori, previous works have shown~\cite{Brewer:2022vkq,Rajagopal:2024lou,Rajagopal:2025nca} that it is often possible to obtain an adiabatic description of the dynamics by introducing suitable time-dependent coordinate redefinitions. In other words, the system is studied in a frame that is adapted to its dynamics.

In particular, Eq.~\eqref{eq: FP Bjorken} was solved in~\cite{Brewer:2022vkq} by introducing arbitrary functions $\AH{A},\AH{B},\AH{G}$ 
that rescale the distribution function and momenta in time: $f(\tau,p_z,p_\perp)= \AH{A}(\tau)\AHf{f}(\tau,\AH{\xi},\AH{\chi})$, with $\AH{\xi}\equiv \AH{G}(\tau)p_z$ and $\AH{\chi}\equiv \AH{B}(\tau)p_\perp$. 
In terms of the rescaled variables, the kinetic equation~\eqref{eq: FP Bjorken} takes the form
\begin{align}
     \partial_\tau \AHf{f}& =-\frac{\AH{A}'(\tau)}{\AH{A}(\tau)}\AHf{f}+\Bigg[\frac{1}{\tau}-\frac{\AH{G}'(\tau)}{\AH{G}(\tau)}\Bigg]\AH{\xi}\partial_{\AH{\xi}}\AHf{f} \nonumber \\
     &\quad +\lambda_0\AH{A}(\tau)^2\AH{G}(\tau)\AHf{I}_{}\partial^2_{\AH{\xi}} \AHf{f} \nonumber\\
     &=-\mathcal{H}\AHf{f} \, .
\label{eq: schrodinger}
\end{align}
For our purposes, we have set $B$ to unity because Eq.~\eqref{eq: FP Bjorken} shows no explicit dependence on $p_\perp$.

The effective Hamiltonian $\mathcal{H}$ is time-dependent through its dependence on $\AH{A}$, $\AH{G}$ and $\AHf{I}_{}=\tfrac{1}{\left(2\pi\right)^3}\int d^2\AH{\chi}\int d\AH{\xi}  \AHf{f}(\tau,\AH{\xi},\AH{\chi})^2$. Additionally, it depends on the state on which it acts through $\AHf{I}_{}$. 
However, the functions $\AH{A}$ and $\AH{G}$ can be chosen such that the eigenstates of $\mathcal{H}$ are time-independent when considered in the rescaled coordinates~\cite{Brewer:2022vkq}. If $\AH{A}$ and $\AH{G}$ satisfy
\begin{align}
    \frac{\AH{G}'(\tau)}{\AH{G}(\tau)} &= \frac{\AH{A}'(\tau)}{\AH{A}(\tau)} +\frac{1}{\tau} \, , \label{eq: AH G-A} \\
    \frac{\AH{A}'(\tau)}{\AH{A}(\tau)} &= -\lambda_0 \AH{A}(\tau)^2\AH{G}(\tau)\AHf{I}_{} \, ,
\label{eq: AH frame}
\end{align}
then the evolution of the system factorizes into the evolution of coordinates and a static eigenmode problem, meaning that it is exactly adiabatic. The eigenvalues $\mathcal{E}_n$ and left and right eigenstates\footnote{ A distinction has to be made between left ($\phi^L$) and right ($\phi^R$) eigenstates because $\mathcal{H}$ is non-Hermitian.} $\phi_n^{L/R}$ of the Hamiltonian $\mathcal{H}$ are
\begin{align}
    \phi_n^L &= \text{He}_{2n}(\AH{\xi}) \, , \\
    \phi_n^R &= \frac{1}{\sqrt{2\pi}(2n)!}\text{He}_{2n}(\AH{\xi})e^{-{\AH{\xi}^2}/{2}} \, ,
    \label{eq: AH eigenvectors}\\
    \mathcal{E}_n &= -2n \frac{\AH{A}'(\tau)}{\AH{A}(\tau)} \, .
\end{align}
The $\text{He}_{2n}$ denote probabilist's Hermite polynomials. 
A general solution to Eq.~\eqref{eq: schrodinger} is given by a linear superposition of the eigenstates $\AHf{f}=\AH{g}(\AH{\chi})\sum_n a_n(\tau) \phi_n^R(\AH{\xi})$, 
where each eigenstate coefficient evolves independently according to $a_n'(\tau) = - \mathcal{E}_n(\tau) a_n(\tau)$, yielding as solutions
\begin{equation}
    a_n(\tau) = a_n(\tau_{\rm ref} ) \left(\frac{\AH{A}(\tau)}{\AH{A}(\tau_{\rm ref})} \right)^{2n} \, .
\end{equation}

In this way, the non-linear partial differential equation~\eqref{eq: FP Bjorken} has been reduced to finding $\AH{A},\AH{G}$ that satisfy Eqs.~\eqref{eq: AH G-A} and~\eqref{eq: AH frame}. We call this pair $\AH{A},\AH{G}$ the \textit{adiabatic frame}. These in turn enter the eigenvalues and eigenstates above. Since $\AHf{I}_{}$ depends on the initial conditions for all coefficients, a general closed form solution to these equations seems beyond reach. However, this can be achieved in special cases, which we now examine.

\noindent \textbf{\emph{Prescaling in Bjorken flow.--}} Long before the discoveries of~\cite{Brewer:2022vkq}, it was realized that Eq.~\eqref{eq: FP Bjorken} admits a self-similar solution that describes the main features of the BMSS fixed point~\cite{Berges:2013fga,Schlichting:2019abc,Berges:2020fwq}. 
However, even in this simpler model, previous works rarely explored how the system relaxes toward this solution starting from arbitrary initial conditions.
Studies in QCD EKT simulations~\cite{Mazeliauskas:2018yef} revealed a \textit{prescaling} regime
\begin{align}
f(\tau,p_z,p_\perp)&=\ps{A}(\tau)f_s(\ps{G}(\tau)p_z,\ps{B}(\tau)p_\perp)\nonumber \\
&=\ps{A}(\tau)f_s(\ps{\xi},\ps{\chi}),
\label{eq: self-similar}
\end{align}
with $\ps{A},\ps{B},\ps{G}$ of a more general form than in Eq.~\eqref{eq: ideal NTFP}.
Recently, \cite{Heller:2023mah} showed that prescaling in non-expanding, isotropic NTFPs can be fully captured by a simple time shift, $t \to t - t_*$ in the isotropic version of Eq.~\eqref{eq: ideal NTFP}, a behavior observed experimentally in a cold-atom system~\cite{Gazo:2023exc}. This reflects the fact that after memory loss, the emerging ``internal clock'' of the NTFP is agnostic to the preceding dynamics. Thus, similar to the adiabatic frame, the prescaling frame is adapted to the dynamics of the most important excitation in the system. The difference is that here this excitation is the late time asymptotics, while for the adiabatic frame it is the instantaneous ground state.

In the End Matter we derive closed form expressions for the functions appearing in the generalized self-similar Ansatz Eq.~\eqref{eq: self-similar}. The results are
\begin{align}
    \ps{B}&=1\,,\\
\ps{G}(\tau)&=\frac{\tau}{\tau_{\rm ref}}\ps{A}(\tau)\,,
\label{eq: A vs G}\\
 \ps{A}(\tau)&=\left(\frac{\tau^2\pm\tau_\ast^2}{\tau_{\rm ref}^2}\right)^{-1/3}\,,\\
    f_s(\ps{\xi},\ps{\chi})&=\ps{g}(\ps{\chi})e^{-\ps{\xi}^2/(2 \kappa)}\,,
\end{align}

\noindent where we defined a reference time $\tau_{\rm ref}$ such that the particle number $n_0$ and transverse energy density $\epsilon_{\perp,0}$ at $\tau_0$ are related to those of the scaling function, $n_s$ and $\epsilon_{\perp,s}$, as $\tau_0 n_0=\tau_{\rm ref} n_s$ and $\tau_0 \epsilon_{\perp ,0}=\tau_{\rm ref}\epsilon_{\perp,s}$.

 The time shift $\tau_*$ in $\ps{A}(\tau)$ emerges naturally as the arbitrary constant of the solution. The constant $\kappa$ depends on $f_s$ through

\begin{equation}
    \kappa=\frac{9 \lambda_0^2}{2^8\pi^5} \ps{I}_{\chi}^2 \tau_{\text{ref}}^2
\label{eq: Ias in fs}
\end{equation}

\noindent with $\ps{I}_{\chi}=\left(\int d^2\ps{\chi}~ \ps{g}(\ps{\chi})^2\right)$.

In the following we will refer to $\ps{A}(\tau)^{-1}{f}(\tau,\ps{\xi},\ps{\chi})$ as the prescaling frame. In this frame, the system becomes effectively static when it is in the prescaling regime.

\noindent \textbf{\emph{Quasinormal Modes of an expanding NTFP.--}} Quasinormal modes provide a physically transparent description of how a system relaxes toward an attractor. They correspond to universal decay channels of small perturbations around the attractor, which evolve independently at late times. In systems exhibiting self-similar dynamics, such as the NTFPs considered here, these modes take the form of power-law corrections, reflecting the absence of intrinsic scales. Each mode is characterized by an exponent $\Omega$ that determines how rapidly it decays, thereby defining a hierarchy of relaxation processes that is independent of the details of the initial state. In \cite{DeLescluze:2025jqx}, it was shown that near a non-expanding, isotropic NTFP, linear-order deviations can be written as a superposition of such modes. We now generalize this construction to the expanding case. 

Consider a perturbation in the prescaling frame $\ps{A}(\tau)^{-1}f(\tau,p_z,p_\perp)=f_s(\ps{\xi},\ps{\chi})+\delta \ps{f}(\tau, \ps{\xi},\ps{\chi})$ with $\delta \ps{f}\ll1$. At linear order in $\delta \ps{f}$, Eq.~\eqref{eq: FP Bjorken} yields an equation of the form
\begin{equation}
    \frac{\ps{A}(\tau)}{\ps{A}'(\tau)}{\partial_\tau}\delta \ps{f}(\tau,\ps{\xi},\ps{\chi})=\hat{O}[f_s](\ps{\xi}) \delta \ps{f}(\tau,\ps{\xi},\ps{\chi}).
    \label{eq: linearized}
\end{equation}

The solutions are the QNMs of the NTFP:
\begin{equation}
    \delta \ps{f}(\tau,\ps{\xi},\ps{\chi})=\delta \cdot \ps{A}(\tau)^\Omega f_{\Omega} (\ps{\xi},\ps{\chi}), \label{eq: QNMs - def}
\end{equation}

\noindent {where $\delta\ll 1$}. Note that in \cite{DeLescluze:2025jqx} the QNMs were defined with $i\Omega$ to be in line with the convention that a negative imaginary part indicates decay. We choose to omit this here to simplify notation.

Similarly to the non-expanding case, the QNM amplitude is a power of the time-dependent amplitude of the prescaling expression, Eq.~\eqref{eq: self-similar}. The similarity to the non-expanding case reflects the fact that, in the prescaling frame, the system behaves as if it was effectively stationary. In this frame, the expansion of the system has been absorbed into a rescaling of variables, and the remaining dynamics corresponds to relaxation toward a fixed point. The QNMs therefore encode the intrinsic relaxation spectrum of this fixed point, independent of the details of the expansion. 

The momentum dependence is determined by an eigenvalue equation,
\begin{multline}
    \Omega f_{\Omega}=\hat{O}[f_s]f_{\Omega}\\=-f_{\Omega}-\ps{\xi} \partial_{\ps{\xi}} f_{\Omega}-\kappa[f_s]\partial^2_{\ps{\xi}} f_{\Omega}-\kappa \frac{\ps{I}_{\delta}[f_s,f_{\Omega}]}{\ps{I}_{s}[f_s]} \partial_{\ps{\xi}}^2 f_s,
\label{eq: eigenvalue equation}
\end{multline}

\noindent where $ \ps{I}_{\delta}=1/(2\pi)^3\int d^2\ps{\chi} \int d\ps{\xi} 2 f_s f_\Omega$. For simplicity, we will in the following assume all perturbations to have the same transverse momentum distribution as the background, and discuss the general case in the Supplemental Material. Eq.~\eqref{eq: eigenvalue equation} can be solved analytically, resulting in the following QNM frequencies and modes. 
\begin{equation}
    \Omega_1=3,\ \Omega_n=2n\ \text{for}\ n=0,2,3,4,...
\label{eq: QNM freqs}
\end{equation}
\begin{equation}
\begin{split}
      &f_{\Omega_1}(\ps{\xi},\ps{\chi})\propto \ps{g}(\ps{\chi}) e^{-{\ps{\xi}^2}/({2\kappa})} \text{He}_{2}(\ps{\xi}/\sqrt{\kappa})\\
      &f_{\Omega_n}(\ps{\xi},\ps{\chi})\propto \ps{g}(\ps{\chi}) e^{-{\ps{\xi}^2}/({2 \kappa})}\Bigg[ \text{He}_{2n}(\ps{\xi}/\sqrt{\kappa})\\
      &\qquad\qquad- \frac{ \ps{I}_{\text{He}_{2n}}}{\sqrt{\pi}\sqrt{\kappa}(\Omega_n-3)}\text{He}_{2}(\ps{\xi}/\sqrt{\kappa})\Bigg],\ n\neq 1.
\end{split}
\label{eq: QNMs}
\end{equation}

\noindent Here, 
$\ps{I}_{ \text{He}_{2n}}=\int d\ps{\xi} e^{-{\ps{\xi}^2}/\kappa}\text{He}_{2n}(\ps{\xi}/\sqrt{\kappa})$ is the longitudinal momentum factor of the integral $\ps{I}$ for the perturbation.

Since both AH and QNM frameworks give rise to the notion of a ground state and excited states that decay away, we investigate how they are related. By comparing the spectra in the AH (Eq.~\eqref{eq: AH eigenvectors}) and the QNM pictures (Eq.~\eqref{eq: QNMs}), we make the following observation: All right AH eigenstates are given by a linear combination of the corresponding QNM and the $\Omega=3$ mode. The latter is related to prescaling as shown in the Supplemental Material. This observation 
gives a first glimpse at how the two frameworks are related.

\noindent \textbf{\emph{Nonlinear solution in terms of QNMs.--}} The QNMs are by definition solutions to the linearized Boltzmann equation in the overoccupied limit, Eq.~\eqref{eq: linearized}. In this section we show how, starting from the QNM solutions, we can construct solutions to the non-linear equation, Eq.~\eqref{eq: FP Bjorken}, that are accurate to arbitrary order in the perturbation. This allows to extend the approximate description of the system's dynamics to earlier times by including higher order deviations from the asymptotic late-time limit. We start by considering Eq.~\eqref{eq: FP Bjorken} in the prescaling frame:
\begin{equation}
    \frac{\ps{A}(\tau)}{\ps{A}'(\tau)}\partial_\tau \psf{f}+\psf{f}+\ps{\xi}\partial_{\ps{\xi}}\psf{f}+\kappa \frac{\psf{I}(\tau)}{\psf{I}_{s}}\partial^2_{\ps{\xi}}\psf{f}=0\,,
    \label{eq: Boltzmann rescaled}
\end{equation}

\noindent where we explicitly denote the time dependence of $\psf{I}(\tau)=\frac{1}{\left(2\pi\right)^3}\int d^2\ps{\chi}\int d\ps{\xi}  \psf{f}^2$ in contrast to the scaling version $\ps{I}_{s}$. Thus far we have just reparameterized our problem, so nothing has changed physically.

To construct a solution of this equation, we start from a linear combination of QNMs on top of the scaling function, which solves it to linear order: $\psf{f}_0=f_s+\sum_n c_n \ps{A}(\tau)^{\Omega_n} f_{\Omega_n}$, where $|c_n|\ps{A}(\tau)^{\Omega_n}\ll1$. A solution to any given order in the perturbation can be found by expanding in the coefficients $c_n \ps{A}(\tau)^{\Omega_n}$. Eq.~\eqref{eq: Boltzmann rescaled} will dictate the appropriate momentum dependence of higher order terms. 
We explain this procedure in more detail in the Supplemental Material.

Here, we give an example. For simplicity we choose $\kappa=1$ by fixing the amplitude of $\ps{g}(\ps{\chi})$, see also Eq.~\eqref{eq: Ias in fs}. Consider only the two lowest QNMs, then the solution to Eq.~\eqref{eq: Boltzmann rescaled} to second order in the $c_n$ is
\begin{equation}
\begin{split}
    \psf{f}&=\ps{g}(\ps{\chi})\Bigg[e^{-\ps{\xi}^2/2}+ c_1 \ps{A}(\tau)^3 e^{-\ps{\xi}^2/2}\left(-1+\ps{\xi}^2\right)\\
    &+c_2 \ps{A}(\tau)^4 e^{-\ps{\xi}^2/2}\left(18-30\ps{\xi}^2+4\ps{\xi}^4\right)\\
    &+\frac{1}{2}c_1^2\ps{A}(\tau)^6e^{-\ps{\xi}^2/2}\left(4-7\ps{\xi}^2+\ps{\xi}^4\right)\\
    &+2 c_1 c_2 \ps{A}(\tau)^7e^{-\ps{\xi}^2/2}\left(-45+114\ps{\xi}^2-33\ps{\xi}^4+2\ps{\xi}^6\right)\\ 
    &+\frac{3}{5} c_2^2 \ps{A}(\tau)^8 e^{-\ps{\xi}^2/2}\left(803-2093\ps{\xi}^2+630\ps{\xi}^4-40\ps{\xi}^6\right)\Bigg].
    \label{eq: nonlinear solution 2 QNMs}
\end{split}
\end{equation}

Similar nonlinear solutions have been considered in the context of black hole QNMs \cite{Carballo:2025ajx,Arnaudo:2025bnm}. There, transient effects are studied, based on the fact that the QNMs are the eigenvalues of a non-normal operator. We anticipate that aspects of these findings could apply in the context of NTFP QNMs.

\noindent \textbf{\emph{Comparison to Adiabatic Hydrodynamization results.--}} 
We now make the physical connection between the AH and QNM descriptions explicit. AH describes the decay of excited states toward an evolving instantaneous ground state, while QNMs describe relaxation directly toward the asymptotic attractor. Therefore, their equivalence depends on whether the combined time dependence of the instantaneous adiabatic frame and decay of excited states in AH can be expressed in terms of QNM decay modes. For this, we need to consider the dynamical regime where the ground state of AH approaches the NTFP. 

Since the gound state of AH matches the prescaling NTFP solution when $\AH{A}(\tau)=\ps{A}(\tau)$, we examine the behavior of $\AH{A}(\tau)$ following Eqs.~\eqref{eq: AH G-A} and~\eqref{eq: AH frame} as it is close to the scaling solution $\ps{A}(\tau)$. No general solution exists but we can construct one perturbatively in the excited states, since these are the source of deviations between $\AH{A}(\tau)$ and $\ps{A}(\tau)$. 

Consider a small perturbation $\propto\phi_n^R$ on top of the right-handed ground state $\phi_0^R$: $\AHf{f}=\phi_0^R(\AH{\xi})+\delta_n \ps{A}(\tau)^{2n}\phi^R_n(\AH{\xi})$, with $\delta_n\ll1$. For $\delta_n=0$, we can match the solution to our prescaling solution, Eqs.~\eqref{eq: NTFP A} and~\eqref{eq: NTFP fs} with $\kappa=1$. To this end, we relate $\AH{A}$ and $\AH{G}$ as in Eq.~\eqref{eq: A vs G} and choose the amplitudes in $\ps{g}(\ps{\chi})$ and $\AH{g}(\AH{\chi})$ such that 

\begin{equation}
\begin{split}
    &\ps{I}_{s}=\frac{2}{3\lambda_0 \tau_{\rm ref}}=\AHf{I}[\AHf{f}=\phi_0^R].
\end{split}
\end{equation}

\noindent Note that this means $\ps{g}(\ps{\chi})=\sqrt{2\pi}~\AH{g}(\AH{\chi})$.

Now, by means of Eqs.~\eqref{eq: AH G-A} and~\eqref{eq: AH frame}, we are able to calculate $\AH{A}$ as a series expansion in $\delta_n$: $\AH{A}=\ps{A}+\sum_{i}^\infty (\delta_n)^i ~\ps{A}_i$. We will consider here the case $n=2$. Plugging the resulting solution into $\AHf{f}=\phi_0^R(\AH{\xi})+\delta_2 \ps{A}(\tau)^{4}\phi^R_2(\AH{\xi})$ and truncating after $O(\delta_2^2)$, we find

\begin{equation}
\begin{split}
    \psf{f}&=\ps{g}(\ps{\chi})\Bigg[e^{-\ps{\xi}^2/2}-(d_1+d_2 \delta_2)\delta_2 \ps{A}(\tau)^3 e^{-\ps{\xi}^2/2}\left(-1+\ps{\xi}^2\right)\\
    &+\frac{1}{96}\delta_2 \ps{A}(\tau)^4 e^{-\ps{\xi}^2/2}\left(18-30\ps{\xi}^2+4\ps{\xi}^4\right)\\
    &+\frac{1}{2}d_1^2\delta_2^2\ps{A}(\tau)^6e^{-\ps{\xi}^2/2}\left(4-7\ps{\xi}^2+\ps{\xi}^4\right)\\
    &-\frac{1}{48} d_1 \delta_2^2 \ps{A}(\tau)^7e^{-\ps{\xi}^2/2}\left(-45+114\ps{\xi}^2-33\ps{\xi}^4+2\ps{\xi}^6\right)\\ 
    &+\frac{1}{15360} \delta_2^2 \ps{A}(\tau)^8 e^{-\ps{\xi}^2/2}\left(803-2093\ps{\xi}^2+630\ps{\xi}^4-40\ps{\xi}^6\right)\Bigg],
    \label{eq: nonlinear solution AH}
\end{split}
\end{equation}

\noindent where we made use of $\AH{A}(\tau)\AHf{f}=f=\ps{A}(\tau)\ps{f}$ and $\AH{G}(\tau)^{-1}\AH{\xi}= p_z=\ps{G}(\tau)^{-1}\ps{\xi}$ together with Eq.~\eqref{eq: AH G-A}. $d_1$ and $d_2$ are integration constants that are picked up order by order in the solution for $\AH{A}$. 

Comparing this to Eq.~\eqref{eq: nonlinear solution 2 QNMs}, we find that there is an exact match when considering terms to $O(\delta_2^2)$ after identifying $c_1=-(d_1+d_2\delta_2)\delta_2$ and $c_2=\delta_2/96$. This serves as a non-trivial cross-check of the earlier solution in Eq.~\eqref{eq: nonlinear solution 2 QNMs}, since this approach is agnostic to the expressions for the QNMs. Therefore, we achieved our goal of understanding the explicit time dependence of $\AH{A}$ close to the NTFP. It is indeed given by a series of power-laws following the QNM picture we presented before. Note that, contrary to what one may have expected from the AH picture, the slowest time evolution is with $\ps{A}^3$ instead of $\ps{A}^2$.

More importantly, we have now proven that the correspondence between QNMs and AH modes we saw at the linear level (see text below Eq.~\eqref{eq: QNMs}) also holds in the perturbative nonlinear regime. The second excited state in AH excites the second QNM $\Omega_2=4$ and the prescaling QNM $\Omega_1=3$. Notably, no contributions from other QNMs enter the solution. This structure persists when considering additional excited states on top of the ground state, where more generally the $n$th excited state can be described by the $n$th and the prescaling QNMs.



\noindent \textbf{\emph{Outlook.--}} While our analysis has focused on a simplified kinetic model, the mechanism underlying the AH–QNM correspondence is more general. It relies on the existence of a self-similar attractor and a well-defined linearized evolution operator around it. Near such attractors, the dominant time dependence can be absorbed into a rescaling of variables, rendering the dynamics effectively stationary and naturally giving rise to an adiabatic description with a hierarchy of modes. In this sense, adiabaticity is not an independent assumption but a consequence of the scaling structure. The nonlinear dynamics can then be organized in terms of QNMs, at least perturbatively.

Our results suggest a route toward constructing adiabatic solutions nonperturbatively by using the QNM spectrum as a basis for organizing excitations near the attractor. This provides a systematic framework for connecting far-from-equilibrium dynamics (AH) to near-attractor behavior (QNMs).

Experimentally, direct access to subdominant excitations is challenging. A full verification of the AH picture at early times may not be possible. However, since QNMs correspond to the leading corrections to the attractor, they may be more accessible in cold-atom systems, as indicated by  observations of prescaling~\cite{Gazo:2023exc} related to a particular QNM as explained in the Supplemental Material. Measuring the hierarchy of decay exponents would provide a direct test of the structure identified here.

Extending this analysis beyond the present simplified model, for instance to full QCD EKT or systems with transverse dynamics, is an important direction for future work. In relation to cold-atom gas experiments, it would be interesting to investigate the correspondence discussed here for models exhibiting dual cascade NTFPs, as in the large-$N$ kinetic theory~\cite{Walz:2017ffj}.

On a broader scope, 
recent insights about the role of QNMs in the context of holography~\cite{Arean:2023ejh,Cownden:2023dam,Withers:2018srf,Grozdanov:2019kge,Arnaudo:2025bnm,Carballo:2025ajx} might synergize with the results of this paper to gain an even deeper understanding of the process. Similarly, important insights might be gained from exploring a connection to singulants~\cite{Heller:2021yjh}, which were introduced as an extension of the notion of QNMs beyond near-equilibrium systems, motivated by the large order behavior of the factorially divergent hydrodynamic gradient expansion.

\noindent \textbf{\emph{Acknowledgements.--}} We would like to thank Thimo Preis for his contributions to the study of prescaling in expanding nonthermal attractors during the early stages of this project. We thank Yi Yin for useful comments on the draft. This project has received funding from the European Research Council (ERC) under the European Union’s Horizon 2020 research and innovation programme (grant number: 101089093 / project acronym: High-TheQ). Views and opinions expressed are however those of the authors only and do not necessarily reflect those of the European Union or the European Research Council. Neither the European Union nor the granting authority can be held responsible for them. This work was partially supported  by the Priority Research Area Digiworld under the program Excellence Initiative  - Research University at the Jagiellonian University in Krakow. The work of BSH was supported in part by grant NSF PHY-2309135 to the Kavli Institute for Theoretical Physics (KITP) and by grant 994312 from the Simons Foundation. 
The work of AM is supported by the DFG through Emmy Noether Programme (project number 496831614) and through CRC 1225 ISOQUANT (project number 27381115).
We thank ECT* for support at the workshop ``Attractors and thermalization in nuclear collisions and cold quantum gases," where we profited from helpful discussions.

\bibliographystyle{bibstyl}
\bibliography{qnm_Ah} 

\providecommand{\href}[2]{#2} \providecommand{\beforedoihref}{} \providecommand{\afterdoihref}{}\begingroup\raggedright\begin{thebibliography}{10}

\bibitem{Berges:2008wm}
J.~Berges, A.~Rothkopf and J.~Schmidt, {\it {Non-thermal fixed points: Effective weak-coupling for strongly correlated systems far from equilibrium}},  \beforedoihref\href{http://dx.doi.org/10.1103/PhysRevLett.101.041603}{Phys. Rev. Lett.}\afterdoihref\  {\bf 101} (2008) 041603 [\href{http://arXiv.org/abs/0803.0131}{{arXiv:0803.0131}}].

\bibitem{Berges:2014bba}
J.~Berges, K.~Boguslavski, S.~Schlichting and R.~Venugopalan, {\it {Universality far from equilibrium: From superfluid Bose gases to heavy-ion collisions}},  \beforedoihref\href{http://dx.doi.org/10.1103/PhysRevLett.114.061601}{Phys. Rev. Lett.}\afterdoihref\  {\bf 114} (2015), no.~6 061601 [\href{http://arXiv.org/abs/1408.1670}{{arXiv:1408.1670}}].

\bibitem{Boguslavski:2018phs}
K.~Boguslavski, {\it {Understanding the dynamics of field theories far from equilibrium}},  \beforedoihref\href{http://dx.doi.org/10.22323/1.336.0136}{PoS}\afterdoihref\  {\bf Confinement2018} (2018) 136 [\href{http://arXiv.org/abs/1811.07171}{{arXiv:1811.07171}}].

\bibitem{Mikheev:2023juq}
A.~N. Mikheev, I.~Siovitz and T.~Gasenzer, {\it {Universal dynamics and non-thermal fixed points in quantum fluids far from equilibrium}},  \beforedoihref\href{http://dx.doi.org/10.1140/epjs/s11734-023-00974-7}{Eur. Phys. J. ST}\afterdoihref\  {\bf 232} (2023), no.~20-22 3393--3415 [\href{http://arXiv.org/abs/2304.12464}{{arXiv:2304.12464}}].

\bibitem{Heller:2015dha}
M.~P. Heller and M.~Spalinski, {\it {Hydrodynamics Beyond the Gradient Expansion: Resurgence and Resummation}},  \beforedoihref\href{http://dx.doi.org/10.1103/PhysRevLett.115.072501}{Phys. Rev. Lett.}\afterdoihref\  {\bf 115} (2015), no.~7 072501 [\href{http://arXiv.org/abs/1503.07514}{{arXiv:1503.07514}}].

\bibitem{Soloviev:2021lhs}
A.~Soloviev, {\it {Hydrodynamic attractors in heavy ion collisions: a review}},  \beforedoihref\href{http://dx.doi.org/10.1140/epjc/s10052-022-10282-4}{Eur. Phys. J. C}\afterdoihref\  {\bf 82} (2022), no.~4 319 [\href{http://arXiv.org/abs/2109.15081}{{arXiv:2109.15081}}].

\bibitem{Jankowski:2023fdz}
J.~Jankowski and M.~Spali{\'n}ski, {\it {Hydrodynamic attractors in ultrarelativistic nuclear collisions}},  \beforedoihref\href{http://dx.doi.org/10.1016/j.ppnp.2023.104048}{Prog. Part. Nucl. Phys.}\afterdoihref\  {\bf 132} (2023) 104048 [\href{http://arXiv.org/abs/2303.09414}{{arXiv:2303.09414}}].

\bibitem{Tanji:2017suk}
N.~Tanji and R.~Venugopalan, {\it {Effective kinetic description of the expanding overoccupied Glasma}},  \beforedoihref\href{http://dx.doi.org/10.1103/PhysRevD.95.094009}{Phys. Rev. D}\afterdoihref\  {\bf 95} (2017), no.~9 094009 [\href{http://arXiv.org/abs/1703.01372}{{arXiv:1703.01372}}].

\bibitem{Chantesana:2018qsb}
I.~Chantesana, A.~Pi{\~n}eiro~Orioli and T.~Gasenzer, {\it {Kinetic theory of nonthermal fixed points in a Bose gas}},  \beforedoihref\href{http://dx.doi.org/10.1103/PhysRevA.99.043620}{Phys. Rev. A}\afterdoihref\  {\bf 99} (2019), no.~4 043620 [\href{http://arXiv.org/abs/1801.09490}{{arXiv:1801.09490}}].

\bibitem{Mazeliauskas:2018yef}
A.~Mazeliauskas and J.~Berges, {\it {Prescaling and far-from-equilibrium hydrodynamics in the quark-gluon plasma}},  \beforedoihref\href{http://dx.doi.org/10.1103/PhysRevLett.122.122301}{Phys. Rev. Lett.}\afterdoihref\  {\bf 122} (2019), no.~12 122301 [\href{http://arXiv.org/abs/1810.10554}{{arXiv:1810.10554}}].

\bibitem{Schmied:2018upn}
C.-M. Schmied, A.~N. Mikheev and T.~Gasenzer, {\it {Prescaling in a far-from-equilibrium Bose gas}},  \beforedoihref\href{http://dx.doi.org/10.1103/PhysRevLett.122.170404}{Phys. Rev. Lett.}\afterdoihref\  {\bf 122} (2019), no.~17 170404 [\href{http://arXiv.org/abs/1807.07514}{{arXiv:1807.07514}}].

\bibitem{Mikheev:2022fdl}
A.~N. Mikheev, A.~Mazeliauskas and J.~Berges, {\it {Stability analysis of nonthermal fixed points in longitudinally expanding kinetic theory}},  \beforedoihref\href{http://dx.doi.org/10.1103/PhysRevD.105.116025}{Phys. Rev. D}\afterdoihref\  {\bf 105} (2022), no.~11 116025 [\href{http://arXiv.org/abs/2203.02299}{{arXiv:2203.02299}}].

\bibitem{Du:2022bel}
X.~Du, M.~P. Heller, S.~Schlichting and V.~Svensson, {\it {Exponential approach to the hydrodynamic attractor in Yang-Mills kinetic theory}},  \beforedoihref\href{http://dx.doi.org/10.1103/PhysRevD.106.014016}{Phys. Rev. D}\afterdoihref\  {\bf 106} (2022), no.~1 014016 [\href{http://arXiv.org/abs/2203.16549}{{arXiv:2203.16549}}].

\bibitem{Heller:2020anv}
M.~P. Heller, R.~Jefferson, M.~Spali{\'n}ski and V.~Svensson, {\it {Hydrodynamic Attractors in Phase Space}},  \beforedoihref\href{http://dx.doi.org/10.1103/PhysRevLett.125.132301}{Phys. Rev. Lett.}\afterdoihref\  {\bf 125} (2020), no.~13 132301 [\href{http://arXiv.org/abs/2003.07368}{{arXiv:2003.07368}}].

\bibitem{Boguslavski:2023jvg}
K.~Boguslavski, A.~Kurkela, T.~Lappi, F.~Lindenbauer and J.~Peuron, {\it {Limiting attractors in heavy-ion collisions}},  \beforedoihref\href{http://dx.doi.org/10.1016/j.physletb.2024.138623}{Phys. Lett. B}\afterdoihref\  {\bf 852} (2024) 138623 [\href{http://arXiv.org/abs/2312.11252}{{arXiv:2312.11252}}].

\bibitem{BarreraCabodevila:2025vir}
S.~Barrera~Cabodevila, X.~Du, C.~A. Salgado and B.~Wu, {\it {Quark production in the bottom-up thermalization}},  \beforedoihref\href{http://dx.doi.org/10.1016/j.physletb.2025.139987}{Phys. Lett. B}\afterdoihref\  {\bf 871} (2025) 139987 [\href{http://arXiv.org/abs/2503.24291}{{arXiv:2503.24291}}].

\bibitem{Brewer:2019oha}
J.~Brewer, L.~Yan and Y.~Yin, {\it {Adiabatic hydrodynamization in rapidly-expanding quark{\textendash}gluon plasma}},  \beforedoihref\href{http://dx.doi.org/10.1016/j.physletb.2021.136189}{Phys. Lett. B}\afterdoihref\  {\bf 816} (2021) 136189 [\href{http://arXiv.org/abs/1910.00021}{{arXiv:1910.00021}}].

\bibitem{Brewer:2022vkq}
J.~Brewer, B.~Scheihing-Hitschfeld and Y.~Yin, {\it {Scaling and adiabaticity in a rapidly expanding gluon plasma}},  \beforedoihref\href{http://dx.doi.org/10.1007/JHEP05(2022)145}{JHEP}\afterdoihref\  {\bf 05} (2022) 145 [\href{http://arXiv.org/abs/2203.02427}{{arXiv:2203.02427}}].

\bibitem{Rajagopal:2024lou}
K.~Rajagopal, B.~Scheihing-Hitschfeld and R.~Steinhorst, {\it {Adiabatic Hydrodynamization and the emergence of attractors: a unified description of hydrodynamization in kinetic theory}},  \beforedoihref\href{http://dx.doi.org/10.1007/JHEP04(2025)028}{JHEP}\afterdoihref\  {\bf 04} (2025) 028 [\href{http://arXiv.org/abs/2405.17545}{{arXiv:2405.17545}}].

\bibitem{Rajagopal:2025nca}
K.~Rajagopal, B.~Scheihing-Hitschfeld and R.~Steinhorst, {\it {Attractors without scaling: adiabatic hydrodynamization with and without inelastic scattering}},  \beforedoihref\href{http://dx.doi.org/10.1007/JHEP03(2026)003}{JHEP}\afterdoihref\  {\bf 03} (2026) 003 [\href{http://arXiv.org/abs/2507.21232}{{arXiv:2507.21232}}].

\bibitem{DeLescluze:2025jqx}
M.~De~Lescluze and M.~P. Heller, {\it {Quasinormal Modes of Nonthermal Fixed Points}},  \beforedoihref\href{http://dx.doi.org/10.1103/tz78-jfbw}{Phys. Rev. Lett.}\afterdoihref\  {\bf 135} (2025), no.~9 091601 [\href{http://arXiv.org/abs/2502.01622}{{arXiv:2502.01622}}].

\bibitem{PineiroOrioli:2015cpb}
A.~Pi{\~n}eiro~Orioli, K.~Boguslavski and J.~Berges, {\it {Universal self-similar dynamics of relativistic and nonrelativistic field theories near nonthermal fixed points}},  \beforedoihref\href{http://dx.doi.org/10.1103/PhysRevD.92.025041}{Phys. Rev. D}\afterdoihref\  {\bf 92} (2015), no.~2 025041 [\href{http://arXiv.org/abs/1503.02498}{{arXiv:1503.02498}}].

\bibitem{Mikheev:2018adp}
A.~N. Mikheev, C.-M. Schmied and T.~Gasenzer, {\it {Low-energy effective theory of nonthermal fixed points in a multicomponent Bose gas}},  \beforedoihref\href{http://dx.doi.org/10.1103/PhysRevA.99.063622}{Phys. Rev. A}\afterdoihref\  {\bf 99} (2019), no.~6 063622 [\href{http://arXiv.org/abs/1807.10228}{{arXiv:1807.10228}}].

\bibitem{Prufer:2018hto}
M.~Pr{\"u}fer, P.~Kunkel, H.~Strobel, S.~Lannig, D.~Linnemann, C.-M. Schmied, J.~Berges, T.~Gasenzer and M.~K. Oberthaler, {\it {Observation of universal dynamics in a spinor Bose gas far from equilibrium}},  \beforedoihref\href{http://dx.doi.org/10.1038/s41586-018-0659-0}{Nature}\afterdoihref\  {\bf 563} (2018), no.~7730 217--220 [\href{http://arXiv.org/abs/1805.11881}{{arXiv:1805.11881}}].

\bibitem{Erne:2018gmz}
S.~Erne, R.~B{\"u}cker, T.~Gasenzer, J.~Berges and J.~Schmiedmayer, {\it {Universal dynamics in an isolated one-dimensional Bose gas far from equilibrium}},  \beforedoihref\href{http://dx.doi.org/10.1038/s41586-018-0667-0}{Nature}\afterdoihref\  {\bf 563} (2018), no.~7730 225--229 [\href{http://arXiv.org/abs/1805.12310}{{arXiv:1805.12310}}].

\bibitem{Glidden:2020qmu}
J.~A.~P. Glidden, C.~Eigen, L.~H. Dogra, T.~A. Hilker, R.~P. Smith and Z.~Hadzibabic, {\it {Bidirectional dynamic scaling in an isolated Bose gas far from equilibrium}},  \beforedoihref\href{http://dx.doi.org/10.1038/s41567-020-01114-x}{Nature Phys.}\afterdoihref\  {\bf 17} (2021), no.~4 457--461 [\href{http://arXiv.org/abs/2006.01118}{{arXiv:2006.01118}}].

\bibitem{Huh:2023xso}
S.~Huh, K.~Mukherjee, K.~Kwon, J.~Seo, J.~Hur, S.~I. Mistakidis, H.~R. Sadeghpour and J.-y. Choi, {\it {Universality class of a spinor Bose{\textendash}Einstein condensate far from equilibrium}},  \beforedoihref\href{http://dx.doi.org/10.1038/s41567-023-02339-2}{Nature Phys.}\afterdoihref\  {\bf 20} (2024), no.~3 402--408 [\href{http://arXiv.org/abs/2303.05230}{{arXiv:2303.05230}}].

\bibitem{Gazo:2023exc}
M.~Gazo, A.~Karailiev, T.~Satoor, C.~Eigen, M.~Ga{\l}ka and Z.~Hadzibabic, {\it {Universal coarsening in a homogeneous two-dimensional Bose gas}},  \beforedoihref\href{http://dx.doi.org/10.1126/science.ado3487}{Science}\afterdoihref\  {\bf 389} (2025), no.~6762 ado3487 [\href{http://arXiv.org/abs/2312.09248}{{arXiv:2312.09248}}].

\bibitem{Lannig:2023fzf}
S.~Lannig, M.~Pr{\"u}fer, Y.~Deller, I.~Siovitz, J.~Dreher, T.~Gasenzer, H.~Strobel and M.~K. Oberthaler, {\it {Observation of two non-thermal fixed points for the same microscopic symmetry}},  6, 2023.

\bibitem{Martirosyan:2023mml}
G.~Martirosyan, C.~J. Ho, J.~Etrych, Y.~Zhang, A.~Cao, Z.~Hadzibabic and C.~Eigen, {\it {Observation of Subdiffusive Dynamic Scaling in a Driven and Disordered Bose Gas}},  \beforedoihref\href{http://dx.doi.org/10.1103/PhysRevLett.132.113401}{Phys. Rev. Lett.}\afterdoihref\  {\bf 132} (2024), no.~11 113401 [\href{http://arXiv.org/abs/2304.06697}{{arXiv:2304.06697}}].

\bibitem{Micha:2002ey}
R.~Micha and I.~I. Tkachev, {\it {Relativistic turbulence: A Long way from preheating to equilibrium}},  \beforedoihref\href{http://dx.doi.org/10.1103/PhysRevLett.90.121301}{Phys. Rev. Lett.}\afterdoihref\  {\bf 90} (2003) 121301 [\href{http://arXiv.org/abs/hep-ph/0210202}{{arXiv:hep-ph/0210202}}].

\bibitem{Schlichting:2012es}
S.~Schlichting, {\it {Turbulent thermalization of weakly coupled non-abelian plasmas}},  \beforedoihref\href{http://dx.doi.org/10.1103/PhysRevD.86.065008}{Phys. Rev. D}\afterdoihref\  {\bf 86} (2012) 065008 [\href{http://arXiv.org/abs/1207.1450}{{arXiv:1207.1450}}].

\bibitem{Kurkela:2012hp}
A.~Kurkela and G.~D. Moore, {\it {UV Cascade in Classical Yang-Mills Theory}},  \beforedoihref\href{http://dx.doi.org/10.1103/PhysRevD.86.056008}{Phys. Rev. D}\afterdoihref\  {\bf 86} (2012) 056008 [\href{http://arXiv.org/abs/1207.1663}{{arXiv:1207.1663}}].

\bibitem{Berges:2013eia}
J.~Berges, K.~Boguslavski, S.~Schlichting and R.~Venugopalan, {\it {Turbulent thermalization process in heavy-ion collisions at ultrarelativistic energies}},  \beforedoihref\href{http://dx.doi.org/10.1103/PhysRevD.89.074011}{Phys. Rev. D}\afterdoihref\  {\bf 89} (2014), no.~7 074011 [\href{http://arXiv.org/abs/1303.5650}{{arXiv:1303.5650}}].

\bibitem{Berges:2013fga}
J.~Berges, K.~Boguslavski, S.~Schlichting and R.~Venugopalan, {\it {Universal attractor in a highly occupied non-Abelian plasma}},  \beforedoihref\href{http://dx.doi.org/10.1103/PhysRevD.89.114007}{Phys. Rev. D}\afterdoihref\  {\bf 89} (2014), no.~11 114007 [\href{http://arXiv.org/abs/1311.3005}{{arXiv:1311.3005}}].

\bibitem{Berges:2013lsa}
J.~Berges, K.~Boguslavski, S.~Schlichting and R.~Venugopalan, {\it {Basin of attraction for turbulent thermalization and the range of validity of classical-statistical simulations}},  \beforedoihref\href{http://dx.doi.org/10.1007/JHEP05(2014)054}{JHEP}\afterdoihref\  {\bf 05} (2014) 054 [\href{http://arXiv.org/abs/1312.5216}{{arXiv:1312.5216}}].

\bibitem{AbraaoYork:2014hbk}
M.~C. Abraao~York, A.~Kurkela, E.~Lu and G.~D. Moore, {\it {UV cascade in classical Yang-Mills theory via kinetic theory}},  \beforedoihref\href{http://dx.doi.org/10.1103/PhysRevD.89.074036}{Phys. Rev. D}\afterdoihref\  {\bf 89} (2014), no.~7 074036 [\href{http://arXiv.org/abs/1401.3751}{{arXiv:1401.3751}}].

\bibitem{Kurkela:2015qoa}
A.~Kurkela and Y.~Zhu, {\it {Isotropization and hydrodynamization in weakly coupled heavy-ion collisions}},  \beforedoihref\href{http://dx.doi.org/10.1103/PhysRevLett.115.182301}{Phys. Rev. Lett.}\afterdoihref\  {\bf 115} (2015), no.~18 182301 [\href{http://arXiv.org/abs/1506.06647}{{arXiv:1506.06647}}].

\bibitem{Schlichting:2019abc}
S.~Schlichting and D.~Teaney, {\it {The First fm/c of Heavy-Ion Collisions}},  \beforedoihref\href{http://dx.doi.org/10.1146/annurev-nucl-101918-023825}{Ann. Rev. Nucl. Part. Sci.}\afterdoihref\  {\bf 69} (2019) 447--476 [\href{http://arXiv.org/abs/1908.02113}{{arXiv:1908.02113}}].

\bibitem{Berges:2020fwq}
J.~Berges, M.~P. Heller, A.~Mazeliauskas and R.~Venugopalan, {\it {QCD thermalization: Ab initio approaches and interdisciplinary connections}},  \beforedoihref\href{http://dx.doi.org/10.1103/RevModPhys.93.035003}{Rev. Mod. Phys.}\afterdoihref\  {\bf 93} (2021), no.~3 035003 [\href{http://arXiv.org/abs/2005.12299}{{arXiv:2005.12299}}].

\bibitem{Mueller:1999pi}
A.~H. Mueller, {\it {The Boltzmann equation for gluons at early times after a heavy ion collision}},  \beforedoihref\href{http://dx.doi.org/10.1016/S0370-2693(00)00084-8}{Phys. Lett. B}\afterdoihref\  {\bf 475} (2000) 220--224 [\href{http://arXiv.org/abs/hep-ph/9909388}{{arXiv:hep-ph/9909388}}].

\bibitem{Blaizot:2013lga}
J.-P. Blaizot, J.~Liao and L.~McLerran, {\it {Gluon Transport Equation in the Small Angle Approximation and the Onset of Bose-Einstein Condensation}},  \beforedoihref\href{http://dx.doi.org/10.1016/j.nuclphysa.2013.10.010}{Nucl. Phys. A}\afterdoihref\  {\bf 920} (2013) 58--77 [\href{http://arXiv.org/abs/1305.2119}{{arXiv:1305.2119}}].

\bibitem{Baier:2000sb}
R.~Baier, A.~H. Mueller, D.~Schiff and D.~T. Son, {\it {'Bottom up' thermalization in heavy ion collisions}},  \beforedoihref\href{http://dx.doi.org/10.1016/S0370-2693(01)00191-5}{Phys. Lett. B}\afterdoihref\  {\bf 502} (2001) 51--58 [\href{http://arXiv.org/abs/hep-ph/0009237}{{arXiv:hep-ph/0009237}}].

\bibitem{Horowitz:1999jd}
G.~T. Horowitz and V.~E. Hubeny, {\it {Quasinormal modes of AdS black holes and the approach to thermal equilibrium}},  \beforedoihref\href{http://dx.doi.org/10.1103/PhysRevD.62.024027}{Phys. Rev. D}\afterdoihref\  {\bf 62} (2000) 024027 [\href{http://arXiv.org/abs/hep-th/9909056}{{arXiv:hep-th/9909056}}].

\bibitem{Kovtun:2005ev}
P.~K. Kovtun and A.~O. Starinets, {\it {Quasinormal modes and holography}},  \beforedoihref\href{http://dx.doi.org/10.1103/PhysRevD.72.086009}{Phys. Rev. D}\afterdoihref\  {\bf 72} (2005) 086009 [\href{http://arXiv.org/abs/hep-th/0506184}{{arXiv:hep-th/0506184}}].

\bibitem{Berti:2009kk}
E.~Berti, V.~Cardoso and A.~O. Starinets, {\it {Quasinormal modes of black holes and black branes}},  \beforedoihref\href{http://dx.doi.org/10.1088/0264-9381/26/16/163001}{Class. Quant. Grav.}\afterdoihref\  {\bf 26} (2009) 163001 [\href{http://arXiv.org/abs/0905.2975}{{arXiv:0905.2975}}].

\bibitem{Brewer:2022ifw}
J.~Brewer, W.~Ke, L.~Yan and Y.~Yin, {\it {Far-from-equilibrium slow modes and momentum anisotropy in an expanding plasma}},  \beforedoihref\href{http://dx.doi.org/10.1103/PhysRevD.109.L091504}{Phys. Rev. D}\afterdoihref\  {\bf 109} (2024), no.~9 L091504 [\href{http://arXiv.org/abs/2212.00820}{{arXiv:2212.00820}}].

\bibitem{Heller:2023mah}
M.~P. Heller, A.~Mazeliauskas and T.~Preis, {\it {Prescaling Relaxation to Nonthermal Attractors}},  \beforedoihref\href{http://dx.doi.org/10.1103/PhysRevLett.132.071602}{Phys. Rev. Lett.}\afterdoihref\  {\bf 132} (2024), no.~7 071602 [\href{http://arXiv.org/abs/2307.07545}{{arXiv:2307.07545}}].

\bibitem{Carballo:2025ajx}
J.~Carballo, C.~Pantelidou and B.~Withers, {\it {Non-modal effects in black hole perturbation theory: transient superradiance}},  \beforedoihref\href{http://dx.doi.org/10.1007/JHEP08(2025)179}{JHEP}\afterdoihref\  {\bf 08} (2025) 179 [\href{http://arXiv.org/abs/2503.05871}{{arXiv:2503.05871}}].

\bibitem{Arnaudo:2025bnm}
P.~Arnaudo, J.~Carballo and B.~Withers, {\it {QNM orthogonality relations for AdS black holes}},  \beforedoihref\href{http://dx.doi.org/10.1007/JHEP09(2025)010}{JHEP}\afterdoihref\  {\bf 09} (2025) 010 [\href{http://arXiv.org/abs/2505.04696}{{arXiv:2505.04696}}].

\bibitem{Walz:2017ffj}
R.~Walz, K.~Boguslavski and J.~Berges, {\it {Large-N kinetic theory for highly occupied systems}},  \beforedoihref\href{http://dx.doi.org/10.1103/PhysRevD.97.116011}{Phys. Rev. D}\afterdoihref\  {\bf 97} (2018), no.~11 116011 [\href{http://arXiv.org/abs/1710.11146}{{arXiv:1710.11146}}].

\bibitem{Arean:2023ejh}
D.~Are\'an, D.~G. Fari\~na and K.~Landsteiner, {\it {Pseudospectra of holographic quasinormal modes}},  \beforedoihref\href{http://dx.doi.org/10.1007/JHEP12(2023)187}{JHEP}\afterdoihref\  {\bf 12} (2023) 187 [\href{http://arXiv.org/abs/2307.08751}{{arXiv:2307.08751}}].

\bibitem{Cownden:2023dam}
B.~Cownden, C.~Pantelidou and M.~Zilh\~ao, {\it {The pseudospectra of black holes in AdS}},  \beforedoihref\href{http://dx.doi.org/10.1007/JHEP05(2024)202}{JHEP}\afterdoihref\  {\bf 05} (2024) 202 [\href{http://arXiv.org/abs/2312.08352}{{arXiv:2312.08352}}].

\bibitem{Withers:2018srf}
B.~Withers, {\it {Short-lived modes from hydrodynamic dispersion relations}},  \beforedoihref\href{http://dx.doi.org/10.1007/JHEP06(2018)059}{JHEP}\afterdoihref\  {\bf 06} (2018) 059 [\href{http://arXiv.org/abs/1803.08058}{{arXiv:1803.08058}}].

\bibitem{Grozdanov:2019kge}
S.~Grozdanov, P.~K. Kovtun, A.~O. Starinets and P.~Tadi\'c, {\it {Convergence of the Gradient Expansion in Hydrodynamics}},  \beforedoihref\href{http://dx.doi.org/10.1103/PhysRevLett.122.251601}{Phys. Rev. Lett.}\afterdoihref\  {\bf 122} (2019), no.~25 251601 [\href{http://arXiv.org/abs/1904.01018}{{arXiv:1904.01018}}].

\bibitem{Heller:2021yjh}
M.~P. Heller, A.~Serantes, M.~Spali{\'n}ski, V.~Svensson and B.~Withers, {\it {Relativistic Hydrodynamics: A Singulant Perspective}},  \beforedoihref\href{http://dx.doi.org/10.1103/PhysRevX.12.041010}{Phys. Rev. X}\afterdoihref\  {\bf 12} (2022), no.~4 041010 [\href{http://arXiv.org/abs/2112.12794}{{arXiv:2112.12794}}].

\bibitem{Preis:2025oar}
T.~Preis, {\em {Strongly correlated quantum fields in and out of equilibrium}}.
\newblock PhD thesis, U. Heidelberg (main), 4, 2025.

\end{thebibliography}\endgroup

\section*{END MATTER: Prescaling in Bjorken flow}

\label{app: end matter}

As explained in the main text, in QCD EKT simulations~\cite{Mazeliauskas:2018yef} it was found that there exists a \textit{prescaling} regime
\begin{align}
f(\tau,p_z,p_\perp)&=\ps{A}(\tau)f_s(\ps{G}(\tau)p_z,\ps{B}(\tau)p_\perp)\nonumber \\
&=\ps{A}(\tau)f_s(\ps{\xi},\ps{\chi}),
\label{eq: self-similar_app}
\end{align}
with $\ps{A},\ps{B},\ps{G}$ of a more general form than in Eq.~\eqref{eq: ideal NTFP}. Motivated by this, in~\cite{Brewer:2022vkq} it was shown that Eq.~\eqref{eq: FP Bjorken} exhibits prescaling if the system is initialized on the ground state $\phi_0^R$ of the spectrum of $\mathcal{H}$. Simultaneously, stability properties of NTFPs were studied in~\cite{Mikheev:2022fdl}. The relevance of prescaling lies in the fact that systems may not reach the ideal NTFP form of Eq.~\eqref{eq: ideal NTFP} within physical timescales~\cite{Gazo:2023exc}. Nevertheless, by allowing generalized rescalings as in Eq.~\eqref{eq: self-similar}, self-similar behavior can still emerge. In heavy-ion collisions, prescaling may play an important role since thermalization occurs on short timescales~\cite{Schlichting:2019abc,Berges:2020fwq}.

The goal of this section is to derive closed form expressions for the functions appearing in the generalized self-similar Ansatz Eq.~\eqref{eq: self-similar_app}. We replicate the steps carried out in~\cite{Preis:2025oar} (see Appendix B.2 therein) and illustrate all of the features we used in this Letter, specifically in performing QNM analysis.

The functions $\ps{A}$, $\ps{B}$, $\ps{G}$ can be related to each other by considering conservation equations. These can be derived from the Boltzmann equation~\eqref{eq: FP Bjorken} by taking appropriate moments.
In particular, we may consider the transverse energy and number density, given by
\begin{align}
    n&=\int\frac{d^3p}{(2\pi)^3} f \, , \\    
    \epsilon_\perp&=\int\frac{d^3p}{(2\pi)^3}p_\perp~ f \, .
\end{align}
Taking these moments of Eq.~\eqref{eq: FP Bjorken}, we immediately find $\partial_\tau(\tau n)=0$ and $\partial_\tau(\tau \epsilon_\perp)=0$, because in both cases the weight of the moment is independent of $p_z$ and the kernel on the RHS contains derivatives in $p_z$, which lead to vanishing boundary terms. Thus, 
\begin{align}
    \tau n &= \tau \ps{A}(\tau)\ps{B}(\tau)^{-2}\ps{G}(\tau)^{-1}n_s=\tau_0n_0,
    \label{eq: n conservation}\\
        \tau \epsilon_\perp&=\tau \ps{A}(\tau)\ps{B}(\tau)^{-3}\ps{G}(\tau)^{-1}\epsilon_{\perp,s}=\tau_0\epsilon_{\perp,0},
\label{eq: e conservation}
\end{align}

\noindent with $n_s$ and $\epsilon_{\perp,s}$ being the corresponding moments of the scaling function $f_s$ in terms of scaled momenta. Here, $\tau_0$, $n_0$, $n_s$, $\epsilon_{\perp,0}$ and $\epsilon_{\perp,s}$ are constants in time. From Eqs.~\eqref{eq: n conservation} and~\eqref{eq: e conservation} we see that $\ps{B}(\tau)$ should be a constant $\ps{B}(\tau)=\frac{n_0 \epsilon_s}{n_s \epsilon_0}\equiv1$ and $A$ and $G$ are related by
\begin{equation}
    \ps{G}(\tau)=\frac{\tau}{\tau_{\rm ref}}\ps{A}(\tau).
    \label{eq: A vs G_app}
\end{equation}
where we defined a reference time $\tau_{\rm ref}$ such that $\tau_0 n_0=\tau_{\rm ref} n_s$ and $\tau_0 \epsilon_{\perp ,0}=\tau_{\rm ref}\epsilon_{\perp,s}$. In this way we give the reference time a physical meaning by relating it to the conserved quantities at $\tau=\tau_0$ and the ones of the scaling function. Note that Eq.~\eqref{eq: A vs G_app} is a solution of Eq.~\eqref{eq: AH G-A} relating $A$ and $G$ in the AH framework.

We can understand Eq.~\eqref{eq: A vs G_app} in the following way. The factor $\tau/\tau_{\text{ref}}$ comes from the expansion term in the left-hand side of Eq.\eqref{eq: FP Bjorken}. The coordinate $(\tau/\tau_{\text{ref}})~p_z$ 
would describe a system that is comoving with expansion. 
This coordinate transformation is compatible with scaling of 
the collision kernel we consider. In this comoving frame, $\ps{A}$ and $\ps{G}$ are simply proportional to each other. Here, we can identify a parallel
to the isotropic case, where $\ps{A}=\ps{G}^\sigma$ \cite{Heller:2023mah}. The exponent $\sigma$ 
is given by the dimensionality $d$ of a system that conserves particle number, while under energy conservation it also depends on the power $z$ of momentum in the dispersion relation of quasiparticles, $\sigma=d+z$.
In our case, the Boltzmann equation is effectively a 1D model that describes only dynamics in the longitudinal direction. As such, the conservation laws, Eqs.~\eqref{eq: n conservation} and~\eqref{eq: e conservation}, are related to the $p_z$-profile only. $d=1$ leads to $\ps{A} \propto \ps{G}$ in the frame comoving with expansion, following the logic in the non-expanding, isotropic case.

The self-similar time evolution Ansatz in Eq.~\eqref{eq: self-similar} implies that $f$ factorizes into a time- and a momentum-dependent part when expressed in terms of the scaled coordinates $\xi$ and $\chi$. Then, the Boltzmann equation~\eqref{eq: FP Bjorken}  for the self-similar Ansatz Eq.~\eqref{eq: self-similar} can be solved by separation of variables. 
\begin{equation}
    \partial_\tau \ps{A}(\tau)=c\ps{A}(\tau)^4\tau.
    \label{eq: A}
\end{equation}
\begin{equation}
     f_s +\ps{\xi}\partial_{\ps{\xi}} f_s= -\mathcal{\kappa}~ \partial^2_{\ps{\xi}}f_s.
\end{equation}

\noindent Here, $c$ is a separation-of-variables constant and $\kappa=-\lambda_0/(\tau_{\text{ref}}~c)\ps{I}_{s}[f_s]$, with $\ps{I}_{s}=1/(2\pi)^3\int d^2\ps{\chi}\int d\ps{\xi} f_s^2$ the effective temperature factor expressed in terms of scaling quantities. The solutions to these equations are as follows: 
\begin{equation}
    \ps{A}(\tau)=\left(\frac{\tau^2\pm\tau_\ast^2}{\tau_{\rm ref}^2}\right)^{-1/3},
\label{eq: NTFP A}
\end{equation}
\begin{equation}
    f_s(\ps{\xi},\ps{\chi})=\ps{g}(\ps{\chi})e^{-\ps{\xi}^2/(2 \kappa)}.
\label{eq: NTFP fs}
\end{equation}

\noindent The time shift $\tau_*$ in $\ps{A}(\tau)$ emerges naturally as the arbitrary constant of the solution. Also, we chose to define $\tau^2_{\rm ref}=-2/(3c)$ to make apparent that for $\tau_*=0$ the solution reduces to the BMSS scaling expression $\left(\tau/\tau_{\rm ref}\right)^{\alpha_\infty}$ with $\alpha_\infty=-2/3$. The equation for $f_s$ technically has two solutions, but only one of them integrates to finite values when computing physical observables. Considering $\kappa$ to be a constant, the equation does not have any $\ps{\chi}$-dependence and $g(\ps{\chi})$ is an arbitrary integrable function. 
However, $\kappa$ depends on $f_s$ through $\ps{I}_{s}$. Consistency implies that
\begin{equation}
    \kappa=\frac{9 \lambda_0^2}{2^8\pi^5} \ps{I}_{\chi}^2 \tau_{\text{ref}}^2
\label{eq: Ias in fs_app}
\end{equation}

\noindent with $\ps{I}_{\chi}=\left(\int d^2\ps{\chi}~ \ps{g}(\ps{\chi})^2\right)$. The equation above shows how $\tau_{\text{ref}}$ enters the scaling function. 

Eqs.~\eqref{eq: A vs G_app}, \eqref{eq: NTFP A} and \eqref{eq: NTFP fs} define the prescaling solution of Eq.~\eqref{eq: FP Bjorken}.
Although beyond our scope, the same ansatz could be applied to prescaling in the full EKT~\cite{Preis:2025oar}, though the scaling-breaking terms add complexity~\cite{Heller:2023mah}.

\newpage
\clearpage
\pagenumbering{arabic}

\appendix
\begin{widetext}

\begin{center}
 {\Large SUPPLEMENTAL MATERIAL}
\end{center}

\section{QNM solution with additional transverse momentum dependence}
\label{app: transverse dependence}

Here we give the solutions to Eq.~\eqref{eq: eigenvalue equation} considering perturbations with transverse momentum dependence that is different than that of the scaling function $\ps{g}(\ps{\chi})$:

\begin{equation}
    \Omega_1=3,\ \Omega_n=2n\ \text{for}\ n=0,2,3,4,...
\end{equation}
\begin{equation}
\begin{split}
      &f_{\Omega_1}(\ps{\xi},\ps{\chi})\propto \ps{g}(\ps{\chi}) e^{-{\ps{\xi}^2}/({2\kappa})} \text{He}_{2}(\ps{\xi}/\sqrt{\kappa})\\
      &f_{\Omega_n}(\ps{\xi},\ps{\chi})\propto e^{-{\ps{\xi}^2}/({2 \kappa})}\Bigg[ \ps{g}_{n}(\ps{\chi})\text{He}_{2n}(\ps{\xi}/\sqrt{\kappa})+\ps{g}(\ps{\chi})\frac{ \ps{I}_{\text{He}_{2n}} \ps{I}_{ \chi n}}{\sqrt{\pi}\sqrt{\kappa} \ps{I}_{s\chi}(\Omega_n-3)}\text{He}_{2}(\ps{\xi}/\sqrt{\kappa})\Bigg],\ n\neq 1.
\end{split}
\label{eq: appendix QNMs}
\end{equation}

\noindent The additional transverse momentum dependence $\ps{g}_{n}(\ps{\chi})$ is an arbitrary function so long as it describes a physical (i.e., positive) distribution function, $\ps{I}_{s\chi}=\int d^2\ps{\chi} \ps{g}(\ps{\chi})^2$ is the transverse momentum factor of the integral $\ps{I}_{s}$, and $\ps{I}_{ \chi n}=\int d^2\ps{\chi} \ps{g}(\ps{\chi}) \ps{g}_{\delta,n}(\ps{\chi})$ is the same integral for the perturbation, with the corresponding longitudinal momentum factor $\ps{I}_{ \text{He}_{2n}}=\int d\ps{\xi} e^{-{\ps{\xi}^2}/\kappa}\text{He}_{2n}(\ps{\xi}/\sqrt{\kappa})$.

\section{Physical interpretation of QNMs} 
\label{app: phys QNM}
The first two modes in Eq.~\eqref{eq: QNM freqs} can be understood in physical terms by making connection with the prescaling notion.
As we will now elaborate, shifting $\tau_{\rm ref}$ and $\tau_*$ lead to the zero mode $\Omega_0=0$ and to $\Omega_1=3$, respectively. We will refer to $\Omega_1=3$ as the prescaling QNM in the following, since a non-zero $\tau_*$ leads to prescaling. We will showcase how the prescaling mode arises under a shift $\tau_* \rightarrow\tau_*+\delta\tau_*$. The effects on the prescaling functions $\ps{A}$ and $\ps{\xi}=\ps{G}(t)p_z$ is as follows [see Eqs.~\eqref{eq: NTFP A} and~\eqref{eq: A vs G}]:
\begin{equation}
\begin{split}
    &\ps{A}(\tau) \xrightarrow{\tau_*\rightarrow\tau_*+\delta\tau_*}\ps{A}(\tau)\pm \alpha_\infty \frac{\tau_*\delta\tau_*}{\tau_{\rm ref}^2}\ps{A}(\tau)^4 \, , \\
    &\ps{\xi}\xrightarrow{\tau_*\rightarrow\tau_*+\delta\tau_*}\ps{\xi}\pm \alpha_\infty \frac{\tau_*\delta\tau_*}{\tau_{\rm ref}^2}\ps{A}(\tau)^3\ps{\xi}.
\end{split}
\end{equation}
Under these changes $\ps{A}(\tau) f_s(\ps{\xi})$ changes as 
\begin{equation}
    \ps{A}(\tau) f_s(\ps{\xi}) \xrightarrow{\tau_*\rightarrow\tau_*+\delta\tau_*} \ps{A}(\tau) f_s (\ps{\xi}) \pm \alpha_\infty \frac{\tau_*\delta\tau_*}{\tau_{\rm ref}^2}\ps{A}(\tau)^4 \left(f_s(\ps{\xi})+\ps{\xi}\partial_{\ps{\xi}} f_s(\ps{\xi})\right) .
\end{equation}

\noindent For our solution of the scaling function Eq.~\eqref{eq: NTFP fs}, this indeed results in the QNM solution with $\Omega_1=3$, see Eq.~\eqref{eq: QNMs}, in the prescaling frame $\bar{A}(\tau)^{-1}f$.

The analogous analysis for $\tau_{\rm ref}\rightarrow\tau_{\rm ref}+\delta \tau_{\rm ref}$ leads to the QNM solution with $\Omega_0=0$. In this case one needs to take into account the explicit dependence of $f_s$ on $\tau_{\rm ref}$. This enters through $\kappa\propto \tau_{\text{ref}}^2$, see Eq.~\eqref{eq: Ias in fs}.

Finally, we show that in order for the prescaling frame to extend to QNMs, $\delta n=0$ and $\delta \epsilon_\perp=0$ need to hold. For the perturbed system, conservation of particle number leads to: 

\begin{equation}
    \tau_0 n_0 = \tau n =\tau \ps{A}(\tau) \ps{B}(\tau)^{-2}\ps{G}(\tau)^{-1}(n_s+\delta n(\tau)) \, .
\label{eq: prescaling frame consistency}
\end{equation}

\noindent Comparing to the unperturbed case, Eq.~\eqref{eq: n conservation}, gives $\delta n(\tau)=0$. The reasoning for transverse energy density $\epsilon_\perp$ is the same. All the QNMs above, except for $\Omega_0=0$, satisfy this constraint. We thus identify the zero mode with perturbations in our conserved quantities. This is consistent with the zero mode arising from a shift in $\tau_{\rm ref}$, since the latter is related to the conserved combination $\tau_0 n _0$ and $\tau_0 \epsilon_{\perp0}$.

 Importantly, the above means the prescaling frame, defined by the expressions Eq.~\eqref{eq: A vs G} and~\eqref{eq: NTFP A}, is solely defined on the level of the scaling function $f_s$. We can thus use the same $\ps{A}$ and $\ps{G}$ when considering the QNMs on top of $f_s$. 

\section{ Scaling perturbation to arbitrary order}\label{app: general perturbation}

Starting from a linear combination of the zeroth and first order result,
\begin{align}
    \psf{f_0}=f_s+\sum_n c_n \ps{A}(\tau)^{\Omega_n} f_{\Omega_n},
\label{eq: appendix QNM ansatz}
\end{align}
we want to construct solutions to the equation
\begin{equation}
    \frac{\ps{A}(\tau)}{\ps{A}'(\tau)}\partial_\tau \ps{f}+\ps{f}+\ps{\xi}\partial_{\ps{\xi}}\ps{f}+\mathcal{\kappa} \frac{\ps{I}}{\ps{I}_{s}}\partial^2_{\ps{\xi}}\ps{f}=0
    \label{eq: appendix Boltzmann rescaled}
\end{equation}
to arbitrary order in the coefficients $c_n$. 

When plugging Eq.~\eqref{eq: appendix QNM ansatz} into Eq.~\eqref{eq: appendix Boltzmann rescaled}, since QNMs solve the equation up to linear order, only terms that are quadratic and cubic in the QNMs remain. These can be countered by higher order terms. By iteration, the equation can be solved order by order in an expansion of products of the different $c_n \ps{A}(\tau)^{\Omega_n}$. In the end, one only has to solve for the remaining momentum dependence. For example, the term $\propto c_1^2$ is proportional to $\ps{A}(\tau)^{2\Omega_1}=\ps{A}(\tau)^6$. By adding a term $c_1^2\ps{A}(\tau)^{2\Omega_1} P(\ps{\xi}) f_s$ to $\psf{f}_0$ we can solve for $P(\ps{\xi})$. $P(\ps{\xi})$ will be a polynomial, as shown below. These steps can be iterated to find higher order corrections. We now lay out how to obtain the solution to arbitrary order in the coefficients $c_n$.

Following the arguments above, we can construct, starting from $f_0$, a solution to the evolution equation via the Ansatz
\begin{align}
    \psf{f}&=\ps{f}_0+\sum_{|\alpha|\ge 2 } \mathbf{c}^\alpha  \ps{A}(\tau)^{\mathbf{\Omega}\cdot\alpha}P_\alpha(\ps{\xi})f_s\\
    &=\sum_{|\alpha|\ge 0 } \mathbf{c}^\alpha  \ps{A}(\tau)^{\mathbf{\Omega}\cdot\alpha}P_\alpha(\ps{\xi})f_s
    \label{eq: nonlinear solution}
\end{align}
using multi-index notation for $\alpha$. The second equality defines $P_0=1$ and the $P_\alpha$ with $\alpha_i=\delta_{in}$ as the corresponding combinations of Hermite polynomials in the $n$-th QNM, respectively. Plugging this into Eq.~\eqref{eq: appendix Boltzmann rescaled}, coefficient comparison in monomials of the $c_n$ yields a set of equations that relate the $P_\alpha(\ps{\xi})$ and allow to compute them order by order. If all $P_\beta$ for $\beta\le\alpha$, $\beta\neq\alpha$ are known, $P_\alpha$ is obtained as the solution of the following integro-differential equation (setting $\kappa=1$):

\begin{align}
    &(\mathbf{\Omega}\cdot\alpha)P_\alpha(\ps{\xi})-\ps{\xi} P_\alpha'(\ps{\xi})+P_\alpha''(\ps{\xi})+2(\ps{\xi}^2-1)\int\frac{d{\ps{\xi}'}}{\sqrt{\pi}}P_\alpha({\ps{\xi}'})f_s^2({\ps{\xi}'})\nonumber\\
    &=\sum_{\substack{0\neq\beta+\gamma\le\alpha\\\beta,\gamma\neq\alpha }} [(1-\ps{\xi}^2)P_{\gamma}(\ps{\xi})+2\ps{\xi} P_\gamma'(\ps{\xi})-P_\gamma''(\ps{\xi})]\int\frac{d{\ps{\xi}'}}{\sqrt{\pi}}P_{\alpha-\beta-\gamma}({\ps{\xi}}')P_\beta({\ps{\xi}'})f_s^2({\ps{\xi}'}).
\end{align}
Assuming further that the $P_\alpha$ are of the form $P_\alpha(\ps{\xi})=\sum_i c_{\alpha,i}\ps{\xi}^{2i}$, a second coefficient comparison yields algebraic equations that allow to compute the $c_{\alpha,i}$:
\begin{align}
    &(\mathbf{\Omega}\cdot\alpha-2i)c_{\alpha,i}+(2i+2)(2i+1)c_{\alpha,i+1}+2(\delta_{i1}-\delta_{i0})\sum_jc_{\alpha,j}\pi^{-1/2}{\Gamma(j+1/2)}\nonumber\\
    &=\sum_{\substack{0\neq\beta+\gamma\le\alpha\\\beta,\gamma\neq\alpha }}[(4i+1)c_{\gamma,i}-c_{\gamma,i-1}-(2i+2)(2i+1)c_{\gamma,i+1}]\sum_{j,k}c_{\alpha-\beta-\gamma}c_\beta\pi^{-1/2}\Gamma(j+k+1/2).
\end{align}

    \newpage
\end{widetext}
\end{document}